\documentclass[twocolumn]{aastex63}
\usepackage{amsmath}

\shorttitle{Disc tearing}
\shortauthors{Raj \& Nixon}
\begin{document}
\title{Disc tearing: implications for black hole accretion and AGN variability}

\author{A.~Raj}
\affiliation{School of Physics and Astronomy, University of Leicester, Leicester, LE1 7RH, UK}
\author[0000-0002-2137-4146]{C.~J.~Nixon}
\affiliation{School of Physics and Astronomy, University of Leicester, Leicester, LE1 7RH, UK}

\email{cjn@leicester.ac.uk}

\begin{abstract}
  Accretion discs around black holes power some of the most luminous objects in the Universe. Discs that are misaligned to the black hole spin can become warped over time by Lense-Thirring precession. Recent work has shown that strongly warped discs can become unstable, causing the disc to break into discrete rings producing a more dynamic and variable accretion flow. In a companion paper, we present numerical simulations of this instability and the resulting dynamics. In this paper, we discuss the implications of this dynamics for accreting black hole systems, with particular focus on the variability of Active Galactic Nuclei (AGN). We discuss the timescales on which variability might manifest, and the impact of the observer orientation with respect to the black hole spin axis. When the disc warp is unstable near the inner edge of the disc, we find quasi periodic behaviour of the inner disc which may explain the recent quasi periodic eruptions observed in, for example, the Seyfert 2 galaxy GSN~069 and in the galactic nucleus of RX J1301.9+2747. These eruptions are thought to be similar to the `heartbeat' modes observed in some X-ray binaries (e.g.\@ GRS~1915+105 and IGR~J17091-3624). When the instability manifests at larger radii in the disc, we find that the central accretion rate can vary on timescales that may be commensurate with, e.g., changing-look AGN. We therefore suggest that some of the variability properties of accreting black hole systems may be explained by the disc being significantly warped, leading to disc tearing.
\end{abstract}


\section{Introduction}
Accreting black holes come in different sizes, from stellar mass black holes in X-ray binaries \citep[e.g.\@][]{Verbunt:1993aa,Remillard:2006aa} to supermassive black holes in the centres of galaxies \citep[e.g.\@][]{Kormendy:2013aa}. In all cases, the accreting matter is in the form of a disc \citep{Pringle:1981aa}. Disc formation is inevitable as the matter possesses angular momentum which it cannot easily shed, whereas the excess energy associated with orbital eccentricity can be readily lost; collisions between gas orbits turn orbital energy into heat which can be radiated away. Once in the form of a disc, additional energy may be extracted from the gas orbits. Disc viscosity, usually ascribed to turbulence in the disc \citep{Shakura:1973aa}, leads to torques which transport angular momentum outwards through the disc allowing most of the matter to spiral slowly inwards. This liberates gravitational potential energy. Accretion on to black holes extracts large amounts of energy per unit mass accreted, and the high accretion rates that are sustainable on to supermassive black holes in galaxy centres makes them the most luminous continually emitting objects in the Universe.

In simple accretion discs, for example those found in dwarf novae, the standard accretion disc model \citep{Shakura:1973aa} provides a good fit to the observed disc spectrum \cite[e.g.\@][]{Hamilton:2007aa} and the thermal-viscous disc instability model provides a good description of the time-dependence \citep[e.g.\@][]{Lasota:2001aa}. However, even in similar types of system, e.g.\@ the nova-like variables, the observed disc emission can deviate from the standard model and requires additional physical effects to be included \citep[see, for example,][]{Nixon:2019ab}. In the dwarf novae and nova-like variables, the central accretor is a white dwarf. It comes as no surprise that accretion discs around black holes are more complex and often times provide more extreme behaviour. In some cases this is due to environmental effects, such as the surrounding matter responsible for creating broad emission lines in AGN spectra and producing the obscuring `torus' \citep{Antonucci:1993aa,Ho:2008aa}, or to additional physical processes such as relativistic jets and outflows that routinely accompany observations of high luminosity black hole accretion at any mass scale \citep{Russell:2010aa,Fabian:2012aa,King:2015ab}. These, and many other processes, imply that application of the standard disc model to the observed spectra of accreting black hole systems is often inadequate.

Recently, \citet[][see also \citealt{Antonucci:2018aa} and references therein]{Lawrence:2018aa} highlighted a potentially more serious issue concerning the timescale of large amplitude luminosity variations in AGN. He points out (and which \citealt{Antonucci:2018aa} argues was already well-known; cf., e.g., the discussion of \citealt{Alloin:1985aa}) that the timescale on which a standard accretion disc is expected to vary---roughly the viscous timescale of the outer disc regions---is orders of magnitude larger than the observed variability timescales, and thus the standard disc model (comprising planar and circular orbits, with a local viscosity) is incompatible with the observed short-timescale, large-amplitude variability. \cite{Lawrence:2018aa} concludes that standard viscous disc theory be abandoned in favour of either (1) an extreme reprocessing scenario in which all the energy emerges from a quasi-point source and is reprocessed by a passive disc or external media, or (2) non-local processes capable of providing significantly shorter timescales by e.g.\@ large scale magnetic stresses. The latter possibility is similar to the ideas put forth by \citet[][see also the discussion in \citealt{Cannizzaro:2020aa}]{King:2004aa}, but these have not yet been applied to AGN disc parameters. The former scenario appears somewhat consistent with the standard disc model, in which the energy dissipation rate per unit area is proportional to $R^{-3}$, and thus the total energy released by the disc at each radius is proportional to $1/R$. This, quite generally, arises as the gravitational potential scales as $1/R$ and it is gravitational potential energy that is the source of the accretion luminosity. Thus, for example, in the standard model (and accounting for the zero torque inner boundary condition) $\approx 3/4$ of the total energy emitted by the disc is emitted from radii between the inner most stable circular orbit, $R_{\rm ISCO}$, and $10R_{\rm ISCO}$.

Re-processing of this energy flux in the outer disc regions is then required to explain the large optical/UV fluxes observed in AGN, as a large emitting area is required. So, combining the large flux and large emitting area, it seems likely that large scale re-processing is necessary\footnote{Heating of the outer disc is also a required addition to the standard disc model in its application to the lightcurves of soft X-ray transients (SXTs), where the outbursts exhibit long exponential decay \citep[][]{King:1998aa}. In all X-ray binaries, it is well-known that reprocessing of inner disc emission is required to explain the optical to X-ray flux ratios \citep{van-Paradijs:1994aa}. In AGN, the peak temperature of a standard accretion disc typically corresponds to rest-frame wavelengths in the extreme ultra-violet, which, if to be seen (at low-redshift) typically has to be reprocessed into the optical/UV. It is worth noting that an irradiated disc is expected to have $T(R) \propto R^{-1/2}$, in contrast to the typical $T(R) \propto R^{-3/4}$, and by modelling the broadband spectra of 23 Seyfert 1 galaxies, \cite{Cheng:2019aa} find that the temperature profiles are best fit with power-law indices ranging from $-1/2$ to $-3/4$. The suggestion that irradiation and reprocessing of central emission plays an important role is not new \citep[see, for example,][]{Collin-Souffrin:1991aa}.}. Such reprocessing of the central flux may occur most readily in the regions where the disc is strongly warped \citep[cf.][]{Natarajan:1998aa}, as the central regions then `see' a larger surface area of the disc. However, these arguments, which focus on the disc spectral properties, have little to say about the appropriateness or not of the standard disc model for the {\it dynamics} of AGN discs. Instead, as noted by \cite{Lawrence:2018aa}, it is the timescales on which the lightcurves vary which most cleanly provides information about the underlying dynamics and thus is most useful in constraining disc models. What the energy arguments (and subsequent reprocessing) do tell us, is that if the central regions can vary (which is of course where the timescales are generally shortest), then we can expect the bulk emission properties to vary with it -- and with lags between wavebands similar to those observed \citep{DeRosa:2015aa}. A similar conclusion was reached by \cite{Sniegowska:2020aa} who argue that the source of the variability lies in instabilities operating in the accretion disc (see also, for example, \citealt{Alloin:1985aa}). Recently \cite{Ricci:2020aa} showed that the changing-look AGN phenomenon can be associated with rapid evolution of the inner disc regions. 

In this paper we discuss the possibility that disc tearing \citep{Nixon:2012ad}, which occurs in warped accretion discs that achieve a sufficient warp amplitude to be unstable to the disc breaking instability \citep{Dogan:2018aa,Dogan:2020aa}, plays a role in producing variable accretion flows around black holes. For discussion of other possible mechanisms we refer the reader to the discussion in \citet[][see also \citealt{Sniegowska:2020aa}]{Cannizzaro:2020aa}. In Section~\ref{disctearing} we provide an overview of the expected dynamics in unstable warped discs. In Section~\ref{timescales} we give the timescales on which the resulting variability can be expected to manifest. In Section~\ref{geometry} we explore the impact of the system orientation on the observable properties. In Section~\ref{discussion} we provide discussion and we conclude in Section~\ref{conclusions}.

\section{Dynamics of disc tearing}
\label{disctearing}
Warped discs have been found through direct and indirect methods in a variety of astrophysical systems. The most direct evidence now exists from spatially resolved observations of protoplanetary discs \citep[e.g.\@][]{Andrews:2020aa}, and recent observations in this area have connected observed disc structures with disc tearing \citep{Kraus:2020aa}. For discs around black holes, the most compelling evidence of disc warps comes from water masers \citep[][see also \citealt{Maloney:1997aa} for discussion]{Miyoshi:1995aa,Greenhill:1995aa,Greenhill:2003aa}. Further evidence is derived from, for example, the long X-ray periods in X-ray binaries \citep[e.g.\@ the models of][]{Wijers:1999aa,Ogilvie:2001ab} and quasi-periodic oscillations that are fit by the Relativistic Precession Model \citep[RPM;][]{Stella:1998aa,Motta:2014aa}. As a specific example, warped disc models are used to explain the time-dependence of the X-ray flux in the X-ray binary Her X-1 \citep{Scott:2000aa,Leahy:2002aa}. In the case of discs around black holes, when the disc is misaligned to the spin plane of a Kerr black hole the disc orbits precess due to the Lense-Thirring effect. The rate of precession depends on the radius of the disc orbits from the black hole, and thus over time the disc acquires a differential twist and thus becomes warped.

The dynamics of a warped disc differ from that of a planar disc primarily due to a resonance between the orbital motion and the radial pressure gradient produced by the warped disc shape (\citealt{Papaloizou:1983aa}; i.e.\@ the midplanes, and thus the regions of high pressure, between two neighbouring rings are misaligned; see e.g.\@ Fig.~10 of \citealt{Lodato:2007aa}). The radial pressure gradient, which oscillates around the orbit, induces epicyclic motion, which in a near-Keplerian disc also oscillates on the orbital timescale. The resonance leads to strong in-plane motions which communicate the disc warp radially and are damped by the disc turbulence. For the case of weak damping (referred to as ``wavelike'' with $\alpha < H/R$) the result is a propagating warp wave \citep{Papaloizou:1995aa,Pringle:1999aa}, while for the case of strong damping (referred to as ``diffusive'' with $\alpha > H/R$) the disc warp evolves following a diffusion equation \citep{Pringle:1992aa,Ogilvie:1999aa}. Here, we will consider the diffusive case as black hole discs are typically expected to be thin and viscous. For a review of warped disc dynamics, see \cite{Nixon:2016aa}.

Numerical simulations of warped discs have revealed that they can break into discrete rings that can subsequently precess effectively independently \citep[see, for example,][]{Nixon:2012ad}.\footnote{\cite{Nixon:2012ad} used Lagrangian (SPH) hydrodynamics \citep[see also][]{Larwood:1996aa,Larwood:1997aa,Nixon:2013ab,Dogan:2015aa,Kraus:2020aa,Raj:2021aa}. This behaviour has also been found in numerical simulations employing Eulerian (grid) hydrodynamics \citep[e.g.\@][]{Fragner:2010aa}, and in simulations that model the effects of magnetic fields \citep{Liska:2020aa}.} The simulation behaviour, which sees rings of matter break off from the rest of the disc in regions where the warp amplitude\footnote{The warp amplitude, $\psi = R\,\left|\partial\boldsymbol{\ell}/\partial R\right|$ where $\boldsymbol{\ell}$ is the unit angular momentum vector pointing normal to the orbital plane at radius $R$, is a dimensionless measure of the strength of the disc warp.} is high, hints at an underlying instability. \citet[][Section~3]{Ogilvie:2000aa} presents a stability analysis of the warped disc equations, finding that they can becomes viscously unstable. \cite{Dogan:2018aa} explore this instability in detail and show that for $\alpha \lesssim 0.2$ there is always a critical warp amplitude (that depends on the value of $\alpha$) at which the disc becomes unstable, and further they connect the instability with the disc tearing behaviour observed in the numerical simulations, arguing that it provides a physical mechanism underpinning the dynamical behaviour. \cite{Dogan:2020aa} further explore the properties of the instability for the case of non-Keplerian rotation that is relevant to discs around black holes.\footnote{It is worth noting that essentially all of the analytical work on the dynamics of warped discs makes use of a Navier-Stokes (local and isotropic) viscosity (but see, for example, Appendix C of \citealt{Nixon:2015ab}). The numerical work generally makes the same assumption by applying a Navier-Stokes viscosity to model the physical viscosity in a disc (but see, for example, \citealt{Liska:2020aa}). \cite{Pringle:1992aa} notes that this may not be right in real discs where the angular momentum transport and dissipation arises through a combination of turbulent and magnetic processes. It has been shown by \citet[][see also \citealt{Ogilvie:2003aa}]{Torkelsson:2000aa} that hydromagnetic turbulence produced by the magnetorotational instability (MRI) leads to the dissipation of motions induced by a warp at a rate that is consistent with an isotropic viscosity. Similarly, \cite{Zhuravlev:2014aa} note that while the viscosity they measure from their GRMHD simulations is anisotropic, they find that ``the effects of anisotropic viscosity on the evolution [of the disc structure] may be rather small''. \cite{Nealon:2016aa} establish that the results of numerical simulations that employ a Navier-Stokes viscosity provide results that are indistinguishable from simulations that model MRI driven turbulence explicitly. So, to the extent that numerical magnetohydrodynamics is capable of modelling the angular momentum transport and energy dissipation in real discs, we can expect the disc models based upon a Navier-Stokes viscosity to provide a reasonable description of the disc dynamics. \cite{King:2007aa} and \cite{Martin:2019aa} have argued (see also the discussion in \citealt{Pjanka:2020aa}) that current MHD models are inadequate for explaining the observed behaviour in accreting systems, particularly with respect to the magnitude of the angular momentum transport and the nature of the energy dissipation, with the latter in numerical MHD models typically controlled by numerical dissipation. Thus our understanding of accretion disc dynamics is not yet so well established that we can be sure that the disc tearing behaviour seen in numerical simulations (now found with both Lagrangian and Eulerian methods, and with viscosity modelled as either Navier-Stokes or explicit MHD turbulence) can occur in real discs. However, all that is physically needed for the disc to be unstable in this manner is that as the disc warp grows, the disc becomes less and less able to hold itself together. More explicitly, that the internal torque attempting to keep the disc locally flat decreases sufficiently with increasing warp amplitude. This seems reasonable to expect from a physical disc, and this has potentially been seen in the protostellar system GW Ori \citep{Kraus:2020aa}.}

The basic reason for instability is that as warp amplitude increases the internal stresses weaken causing the disc to become less able to hold itself together \citep{Ogilvie:1999aa,Nixon:2012aa}. In a companion paper \citep{Raj:2021aa} we present numerical simulations of discs with various parameters including the disc thickness, viscosity parameter and inclination with respect to the black hole spin. We found that the properties of the simulated discs were consistent with, for example, the predictions for the critical warp amplitude at which the disc becomes unstable. We showed that discs with a lower viscosity parameter become unstable at lower warp amplitudes, and that discs which are thinner and have a larger misalignment with respect to the black hole spin generally reach larger warp amplitudes and are thus more likely to become unstable. Once a disc becomes unstable it breaks into two or more discrete rings, joined by a low-density transition region in which the orbital plane varies sharply with radius. As the warp is produced by the gravitational potential, which causes the orbits to precess, each of the individual rings subsequently precess with only weak interaction with neighbouring rings. However, over time the misalignment between neighbouring rings grows and thus their velocity fields become partially opposed. Thus any subsequent spreading of the rings, or perturbations to the orbits, results in collisions of the gas. This promotes shocks and the loss of rotational support for the gas\footnote{It may appear that angular momentum is not conserved as the gas has experienced a reduction in angular momentum. However, the total angular momentum, including the angular momentum transferred to the black hole through the back reaction of the Lense-Thirring effect, is conserved through this process. Essentially, the disc orbits have borrowed angular momentum from the hole in order to change their orbital plane and thus allow internal cancellations within the disc.}. This puts some of the gas on eccentric orbits (with apocentre at the radius of the original ring), which upon reaching pericentre have either accreted on to the central object, or collide with other gas orbits to circularise at a smaller radius. As the precession timescale must be shorter than the standard viscous timescale in the disc to produce a substantial warp \citep{Nixon:2012ad}, this process enhances the accretion rate on to the central object by delivering matter to small radii faster than viscous torques could do so. The efficiency of this process is determined by the misalignment angle between the disc and the black hole. The maximum angle created between the disc orbits is twice the angle to the black hole spin vector. For small angles, the accretion rate may be enhanced by a factor of order unity. While for larger angles the gas can fall a considerable distance in radius and increase the accretion rate substantially.

As two neighbouring rings precess their mutual misalignment angle oscillates between zero and twice their initial angle to the black hole spin vector\footnote{The maximum angle is somewhat reduced by the rings also slowly aligning with the black hole spin vector, and in general simulations show that the inner ring is closer to alignment than the outer ring. Thus, when the disc is close to the line of stability, i.e.\@ the warp amplitude is not far from the critical value for instability and the growth rates of the instability are small, the internal angle is much less than the maximal value. However, for regions of strong instability and growth rates of order the dynamical timescale, the angle is close to the maximal one.}. Collisions between matter in the rings leads to shocks as the orbital velocity is significantly larger than the sound speed for a disc. The resulting dynamics depends on the local cooling rate of the shocked gas. The relevant timescales are the expansion timescale $\sim H/c_{\rm s}$ (where here the sound speed $c_{\rm s}$ is that of the shocked gas) and the (typically free-free) cooling timescale \citep{Pringle:1979aa}. If it can cool efficiently, then the cooling gas can fall inwards to the new circularisation radius (as above). However, if the cooling is inefficient, as might be expected in the innermost regions of the disc or if the shocks involve only a small amount of matter, the shocked gas takes on a more quasi-spherical distribution and can be supported somewhat against infall by pressure \citep[cf.][]{Stone:1999aa}. This shocked gas may resemble the X-ray corona \citep[as suggested by][]{Nixon:2014aa}, and its subsequent dynamics is yet to be explored fully but may resemble a radiatively inefficient accretion flow \citep[see, for example,][]{Stone:1999aa,Inayoshi:2018aa}.

An alternative possibility is that in these central regions the energy dissipated in the strong warping of the disc is not readily radiated away, and thus on the timescales on which the viscous instability of the warp grows the disc structure may be changed; specifically the disc thickness may increase due to the additional dissipation of energy. This may have a stabilising effect on the viscous instability, but may also lead to thermal instability and would also have a significant impact on the central accretion rate due to the dependence of the viscous timescale on disc temperature. We plan to explore this dynamics in subsequent work.

\section{Timescales}
\label{timescales}
In this section we provide a discussion of the relevant timescales on which we can expect the disc to show variability. We limit the discussion to AGN discs. For discussion of the types of variability and the relevant timescales from disc warping and disc tearing in X-ray binaries we refer to \cite{Nixon:2014aa}. Variability may also be produced by other mechanisms \citep[e.g.\@][]{Sniegowska:2020aa}, and the timescales may be affected by the inclusion of additional physics \citep[e.g.\@][]{Dexter:2019aa}, but here we employ what we consider to be standard AGN disc structure and consider the effect of disc warping alone (for a broad discussion of several other mechanisms we refer the reader to Section 4 of \citealt{Cannizzaro:2020aa}).

AGN discs are expected to be limited in radial extent by their own self-gravity  to a size of order $R_{\rm sg} = 0.02$\,pc, almost independent of disc or black hole parameters \citep{Goodman:2003aa,Levin:2007aa,King:2007ab}. Discs that form or grow beyond this radius are typically gravitationally unstable, with Toomre $Q = c_{\rm s}\kappa/\pi G\Sigma < 1$ \citep{Toomre:1964aa}. This, coupled with rapid cooling rates compared to the local orbital period, results in fragmentation of the disc into stars at radii $R > R_{\rm sg}$ \citep{Gammie:2001aa}. This is consistent with the disc of stars observed in our own galaxy centre which orbit just outside this radius \citep{Genzel:2003aa,Ghez:2005aa}. For sufficiently large SMBH, with $M \gtrsim 5\times 10^{10} M_\odot$, the self-gravity radius can be of the order of the gravitational radius of the black hole and thus the formation of a stable accretion disc is likely to be prohibited \cite[][see also \citealt{Natarajan:2009aa}]{King:2016ab}. For black holes below this limit, but with high mass, say $M \gg 10^8 M_\odot$, the disc extent is limited to 10s to 100s of gravitational radii, and thus the entire disc may be in the regime where radiation pressure provides the dominant vertical support, i.e.\@ determines the disc scale height\footnote{The actual structure of the radiation pressure dominated regions is uncertain, as the standard disc model including a Shakura-Sunyaev viscosity is unstable there \citep{Lightman:1974aa,Shakura:1976aa}.}. While for typical black hole masses of $10^6-10^8 M_\odot$, the self gravity radius is large enough that radiation pressure there is not important.

The basic timescale on which an accretion disc evolves is the viscous timescale on which angular momentum is transported and matter diffuses through the disc, given by \citep{Pringle:1981aa}
\begin{equation}
  \label{tnu1}
  t_\nu = \frac{R^2}{\nu}
\end{equation}
where $R$ is the disc radius, and $\nu$ the kinematic viscosity. The viscosity in an accretion disc is usually modelled as $\nu = \alpha c_{\rm s} H$ \citep{Shakura:1973aa}, and is assumed to originate from hydromagnetic turbulence \citep[see e.g.\@][for discussion]{Martin:2019aa}. In black hole accretion discs outside of quiescence, the accretion rate is typically high enough that the disc is sufficiently ionised for the magnetorotational instability to provide a source of disc turbulence \citep{Balbus:1991aa}. Application of the disc structure presented by \cite{Collin-Souffrin:1990aa}, for the case where radiation pressure is not important, yields \citep{King:2007ab}
\begin{eqnarray}
  \label{tnu2}
  t_\nu(R_{\rm sg}) & = & 1.4\times 10^6 \left(\frac{\alpha}{0.1}\right)^{-2/27}\left(\frac{\epsilon}{0.1}\right)^{22/27} \\
  && \times\left(\frac{L}{0.1L_{\rm Edd}}\right)^{-22/27}\left(\frac{M}{10^8M_\odot}\right)^{-4/27}\,{\rm yr}\,, \nonumber 
\end{eqnarray}
where $\epsilon$ is the radiative efficiency of accretion and $L_{\rm Edd}$ is the Eddington luminosity. Thus the timescale of variability associated with the accretion timescale of the disc itself is long, and commensurate with the timescales on which AGN global properties appear to `flicker' \citep{Schawinski:2015aa,King:2015aa}. However, this timescale is not consistent with short-timescale, large-amplitude variability observed in AGN lightcurves \citep[as has been pointed out by, for example,][]{Lawrence:2018aa}. If this short-timescale variability arises due to viscous inflow of matter, i.e.\@ the luminosity variations are produced by the standard energy dissipation in an accretion disc due to viscous torques, then some mechanism is required for creating time-dependent fluctuations in the mass flow rate at smaller radii.

For smaller radii, we need to account for the effects of radiation pressure on the disc vertical structure \citep{Shakura:1973aa,Novikov:1973aa}. In this case, the viscous timescale can be written as \citep[e.g.\@][]{Dexter:2019aa}
\begin{equation}
  \label{tnu3}
  t_\nu \approx 10 \left(\frac{\alpha}{0.1}\right)^{-1}\left(\frac{M}{10^7 M_\odot}\right)\left(\frac{{\dot m}}{0.1}\right)^{-2}\left(\frac{R}{50R_{\rm g}}\right)^{7/2}\,{\rm yr}\,,
\end{equation}
where ${\dot m}$ is the accretion rate in Eddington units, and we have assumed electron scattering opacity. From this we can see that if mass can be fed in a time-dependent manner to radii of order $10-50R_{\rm g}$, then the corresponding viscous timescales (of weeks to years) are capable of providing the necessary timescales to generate the observed luminosity variations.

For the disc tearing scenario described in Section~\ref{disctearing} above, the other basic timescale is the time taken for disc orbits that are misaligned to the black hole spin to precess around the black hole spin vector -- the nodal (or Lense-Thirring) precession timescale. For a test particle orbiting the black hole, this is given by\footnote{Note that close to the black hole this formula is strictly only valid for small black hole spin. For example, for $a \approx 0.5$ this formula underpredicts (compared to the full Kerr solution) the precession timescale by roughly 20 per cent near the ISCO. Accurate formulae for the frequencies are required for e.g.\@ the Relativistic Precession Model (RPM) of \cite{Stella:1998aa}, and are provided by, for example, \cite{Motta:2014aa}. The RPM requires disc matter to perform orbits that are close to test particle orbits and have non-zero misalignment and eccentricity near the ISCO. The latter requirement, coupled with the success of the RPM in fitting QPOs in X-ray binaries, provides motivation for warped discs to exist at small radii around black holes (and the eccentricity may be provided by interacting rings and the partial exchange of orbital angular momentum between them). While the former requirement, that the orbits behave individually rather than as a collective fluid, requires---as noted by \cite{Dogan:2020aa}---that the disc be unstable to breaking up into discrete parts, which could be a result of disc tearing.}
\begin{eqnarray}
  \label{tp}
  t_{\rm np}(R) & = & \frac{1}{\Omega_{\rm np}} = \frac{c^2R^3}{2GJ_{\rm h}} = \frac{1}{2a}\left(\frac{R}{R_{\rm g}}\right)^3\frac{GM}{c^3} \\
  & = & 0.57 \left(\frac{a}{0.5}\right)^{-1}\left(\frac{R}{10R_{\rm g}}\right)^{3}\left(\frac{M}{10^7M_\odot}\right)\,{\rm days}\, \nonumber
\end{eqnarray}
where $\Omega_{\rm np}$ is the (nodal) precession frequency, $R$ is the disc radius, $J_{\rm h}$ is the black hole angular momentum, $a$ the black hole spin, and $R_{\rm g} = GM/c^2$ the gravitational radius of the black hole.

Equation~\ref{tp} gives the timescale on which a single narrow ring of gas orbiting a black hole undergoes nodal precession. However, in general, precessing rings of gas are accompanied by other precessing rings or a warped and precessing outer disc. For two freely-precessing rings separated by a radial distance of $\Delta R$, the differential precession timescale on which the rings precess apart is given by
\begin{equation}
  \label{dtp}
  \delta t_{\rm np}(R) = \frac{1}{\Omega_{\rm np}(R)-\Omega_{\rm np}(R+\Delta R)} \approx \frac{R}{3\Delta R}t_{\rm np}(R)\,,
\end{equation}
where in the last step we have assumed that $\Delta R \ll R$.~\footnote{We note that if $\Delta R \sim R$, then this patch of the disc no longer follows the precession timescale of a particular disc location, but instead follows the angular momentum weighted average precession timescale \citep[e.g.\@][]{Larwood:1997aa}.} For rings produced by disc tearing, we expect $\Delta R \gtrsim H$ when the disc is strongly unstable and can have $\Delta R \gg H$ when it is only weakly unstable.

As discussed above, the disc dynamics resulting from the instability of the disc warp can produce different scenarios. If the innermost regions of the disc are unstable, then we expect the variability to manifest on the precession timescale of the inner disc regions. In particular, in this case we expect variability on the differential precession timescale (\ref{dtp}), and the disc matter may plunge directly into the hole. If, instead, the disc is unstable further out such that the mass flow from the unstable region does not reach the ISCO, then the matter circularises at a new smaller radius\footnote{Typically, the new orbits here are close to alignment with the black hole spin as the reduction in angular momentum of the material comes from the misaligned component of the original orbits. However, a small residual misalignment is possible and the orbits are initially eccentric. Note that the system as a whole conserves angular momentum, but each ring of gas in the disc borrows angular momentum from the black hole via the Lense-Thirring effect in order to be misaligned to neighbouring rings. Subsequent interaction between the misaligned rings can reduce the orbital angular momentum of both \citep[cf. Eqn~5 of][]{Hall:2014aa}.}. In this case there are several competing timescales for the variability. The (differential) precession timescale determines the timescale for shocks between rings and the inflow of matter. The orbits circularise at their new smaller radius efficiently on the local orbital time (due to additional shocks at smaller radii; cf. \citealt{Nixon:2012ab}). Then, once the material is circularised, a strong increase in disc luminosity follows over a timescale of order the viscous timescale (\ref{tnu3}) from the new radius. It is worth noting that for a ring of material placed at some radius around a central accretor, the luminosity rises rapidly to a peak and then declines at a slower rate \citep[e.g.\@ Fig.~3 of][]{Nixon:2021aa}, which is consistent with the shapes of the lightcurves of some variable AGN (e.g.\@ as noted for J0225+0030 by \citealt{Macleod:2016aa}). An additional source of variability may arise from the changing geometry of the system with respect to the line-of-sight, which we discuss further in Section~\ref{geometry} below.

In Fig.~\ref{Fig1} we provide some example accretion rate curves from the simulations presented in \cite{Raj:2021aa}. In this Figure the accretion rate is plotted on a log scale and in arbitrary units. The time axis has been scaled to days for a $10^7M_\odot$ black hole, which can be rescaled to any mass black hole by multiplying the time axis by $M/10^7M_\odot$. The black line corresponds to the accretion rate in the simulation with $\alpha=0.03$, $H/R = 0.02$, and $\theta = 10^\circ$, where $\theta$ is the initial angle between the disc and black hole spin. The red line corresponds to the accretion rate for the same parameters, except for this time $\theta=60^\circ$. For the $10^\circ$ case, the disc attains a mild warp which does not create a noticeable impact on the disc structure or evolution and thus the accretion rate closely resembles the accretion for a planar disc. In the Figure we see the accretion rate slowly decline with time as the simulated disc loses mass through the inner boundary at $R_{\rm ISCO}$ and spreads to larger radii. In contrast, for the $60^\circ$ case the disc undergoes disc tearing in the inner regions at early times (the first $\approx 10$\,days of evolution), and over time the unstable region moves outwards. After $t\approx 10$\,days, precessing rings at larger radii (10s of $R_{\rm g}$) feed matter into an aligned disc at radii between $R_{\rm ISCO}$ and a few $R_{\rm ISCO}$. Thus the variability in the accretion rate becomes both longer-timescale and lower-amplitude as the innermost aligned disc's viscous timescale regulates the accretion. This shows that short-timescale (weeks to months) and large-amplitude (factors of several) variability in the accretion rate on to supermassive black holes could be caused by disc tearing. We expect that this variability would be reflected in the time-dependent emission of such systems as the accretion rate is a reasonable proxy for the rate at which energy can be extracted from the disc orbits.

\begin{figure}
  \includegraphics[width=\columnwidth]{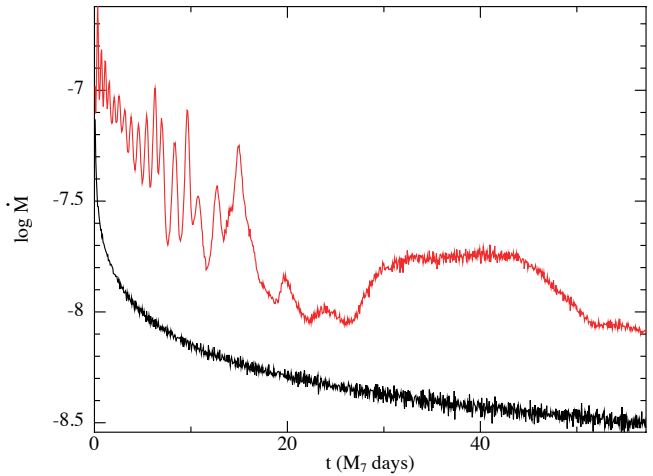}
  \caption{Example accretion rates versus time from the simulations presented in \cite{Raj:2021aa}. The two curves represent the accretion rate in the simulations with $\alpha=0.03$, $H/R=0.02$ and $\theta=10^\circ$ (black line) and $\theta=60^\circ$ (red line). The accretion rate is plotted on a log scale, and is in arbitrary units. The time axis has been scaled to days for a $10^7M_\odot$ black hole, which can be rescaled to any mass by multiplying the time axis by $M/10^7M_\odot$. The $10^\circ$ simulation achieves a mild warp which does not strongly impact the disc structure, and thus the accretion rate is similar to a planar disc with the slow decline caused by the disc accretion and spreading to larger radii with time (the simulations did not model a steady disc with mass input over time). The $60^\circ$ simulation achieves a strong warp which results in disc tearing. At early times the inner disc is unstable, resulting in periodic accretion of rings of matter. At later times the unstable region moves to larger radii, and the inner disc is aligned to the black hole spin. This results in lower-amplitude and longer-timescale variability. Note that the properties of the disc tearing events, e.g.\@ the radius and thus timescale at which they occur, depends on the disc properties and also the black hole spin. Therefore this plot should be taken as representative of the kind of behaviour that may occur, and not a prediction of an accretion rate for any particular system.}
  \label{Fig1}
\end{figure}

\section{Impact of system orientation}
\label{geometry}
We have discussed in the previous section the timescales on which the disc structure is expected to vary, through either accretion (equation~\ref{tnu1}) or precession (equations~\ref{tp} \& \ref{dtp}). In this section we discuss the impact of the orientation of the system with respect to an observer.

In general, the angle between the black hole spin and the accretion disc angular momentum is expected to decrease over time through dissipation between neighbouring rings\footnote{However, there are exceptions to this. For example, if one includes the effects of radiation warping, it is possible for the disc to achieve a shape in which the entire sky---as seen from the black hole---is covered by the disc surface \citep{Pringle:1997aa}. Alternatively, if the total disc angular momentum dominates the angular momentum of the black hole and the disc is initially closer to counteralignment (i.e.\@ the disc-black hole angle is $\theta > 90^\circ$) then while the inner disc regions initially counteralign ($\theta \rightarrow 180^\circ$) with the black hole spin, the outer regions (on a longer timescale) align ($\theta \rightarrow 0^\circ$) with the black hole spin \citep{King:2005aa,Lodato:2006aa} and thus any line of sight may become blocked over time.}; Lense-Thirring precession of each ring maintains the inclination angle with respect to the black hole, and subsequent interaction between neighbouring rings that occupy different planes (e.g.\@ ones that have precessed apart) reduces the inclination to the black hole spin. This means that a line-of-sight that is closer to the black hole spin axis than the original disc misalignment always has a clear view of the disc centre. In this case, any observed variability is caused by either emission from the shocks occurring between precessing rings in unstable regions of the disc or from the variable accretion rate caused by the time-dependent mass flow rates from the unstable regions. 

\begin{figure*}
  \centering
\includegraphics[width=0.161\textwidth]{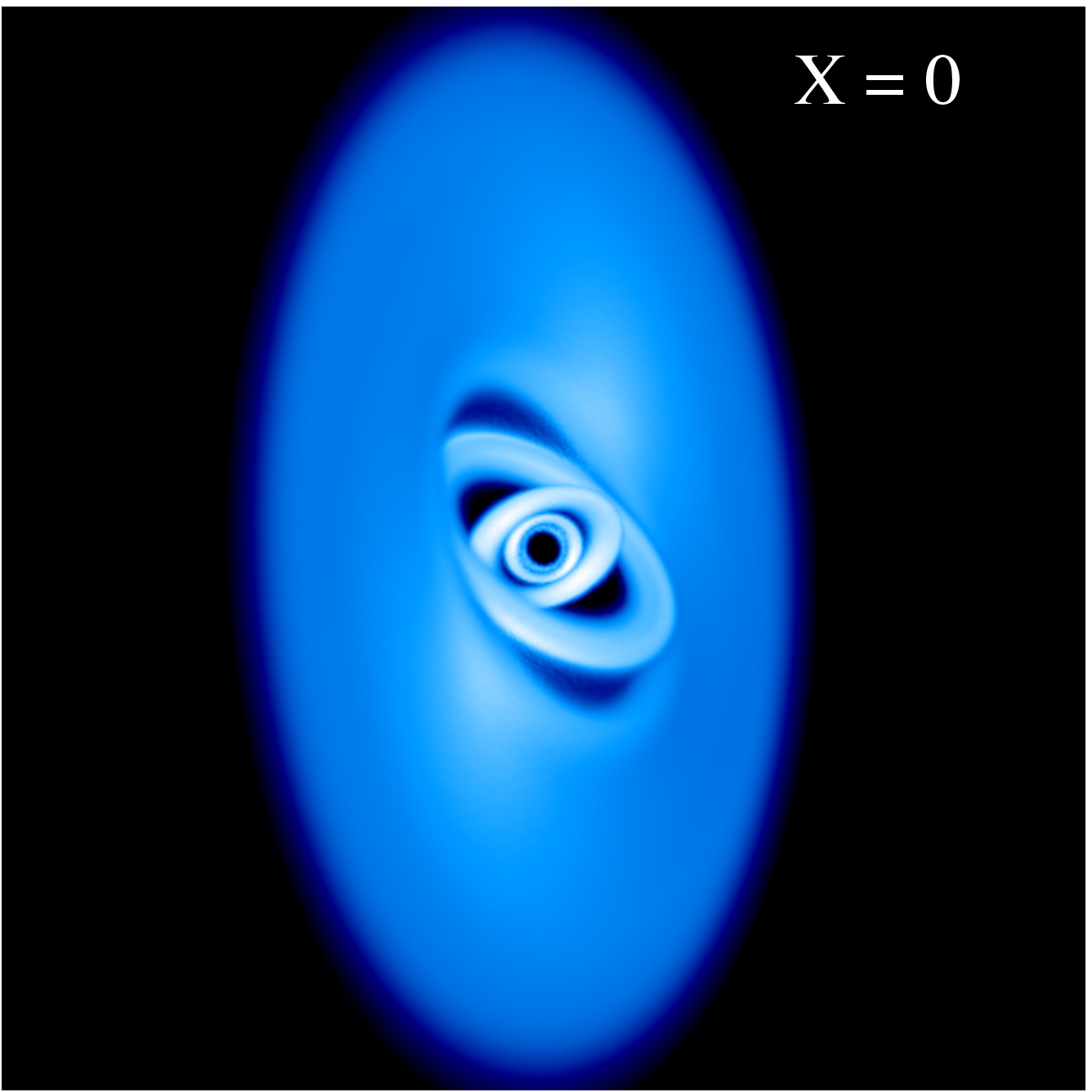}
\includegraphics[width=0.161\textwidth]{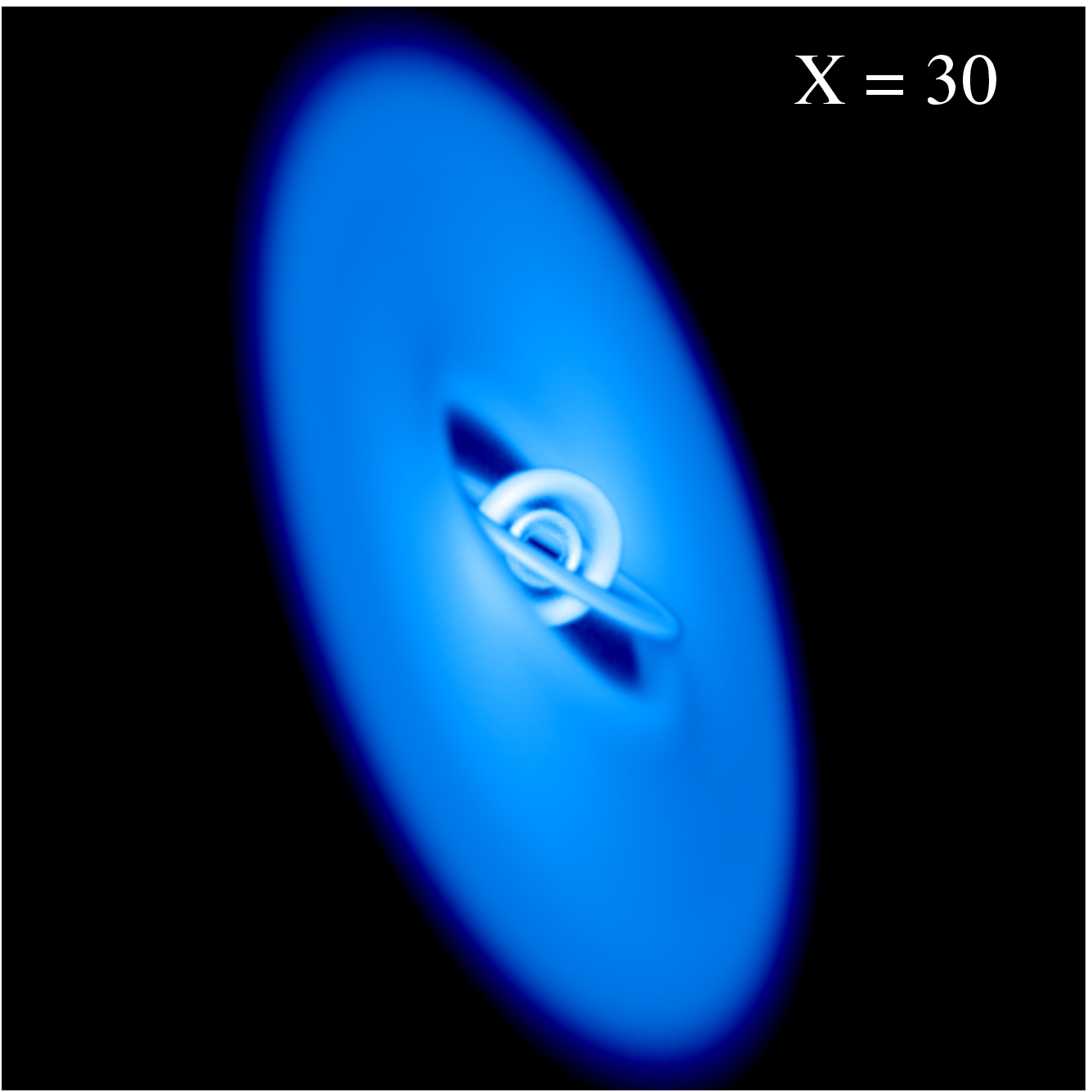}
\includegraphics[width=0.161\textwidth]{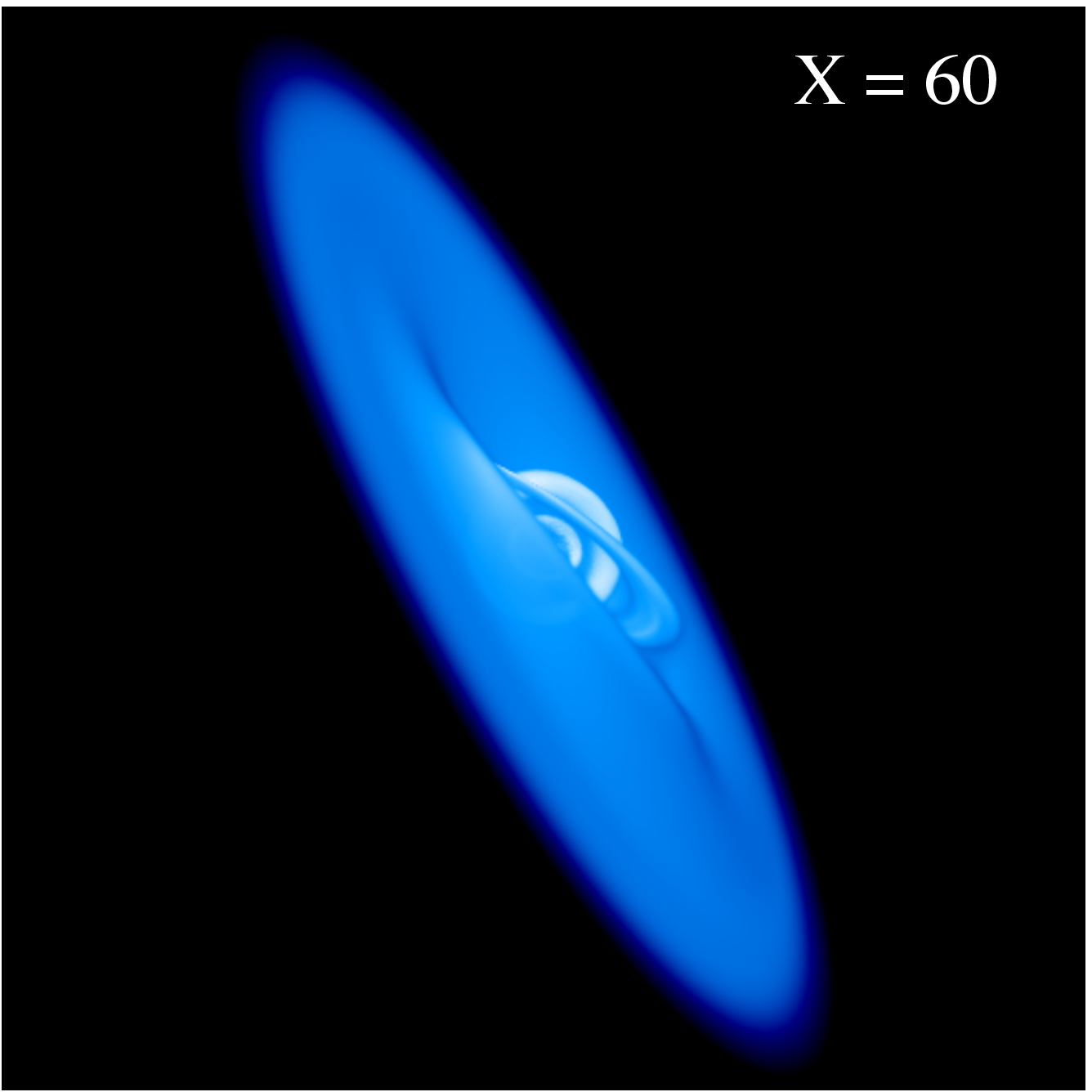}
\includegraphics[width=0.161\textwidth]{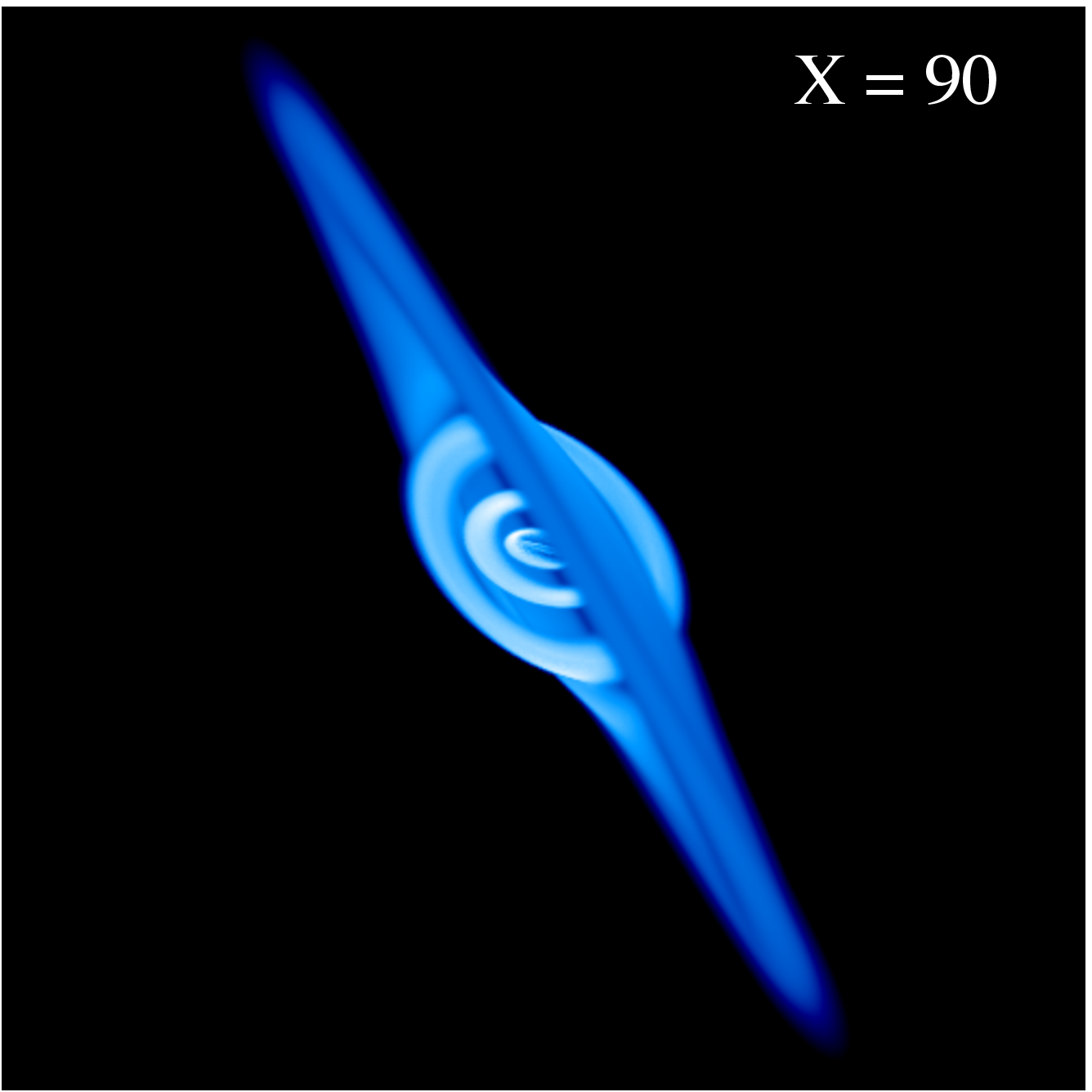}
\includegraphics[width=0.161\textwidth]{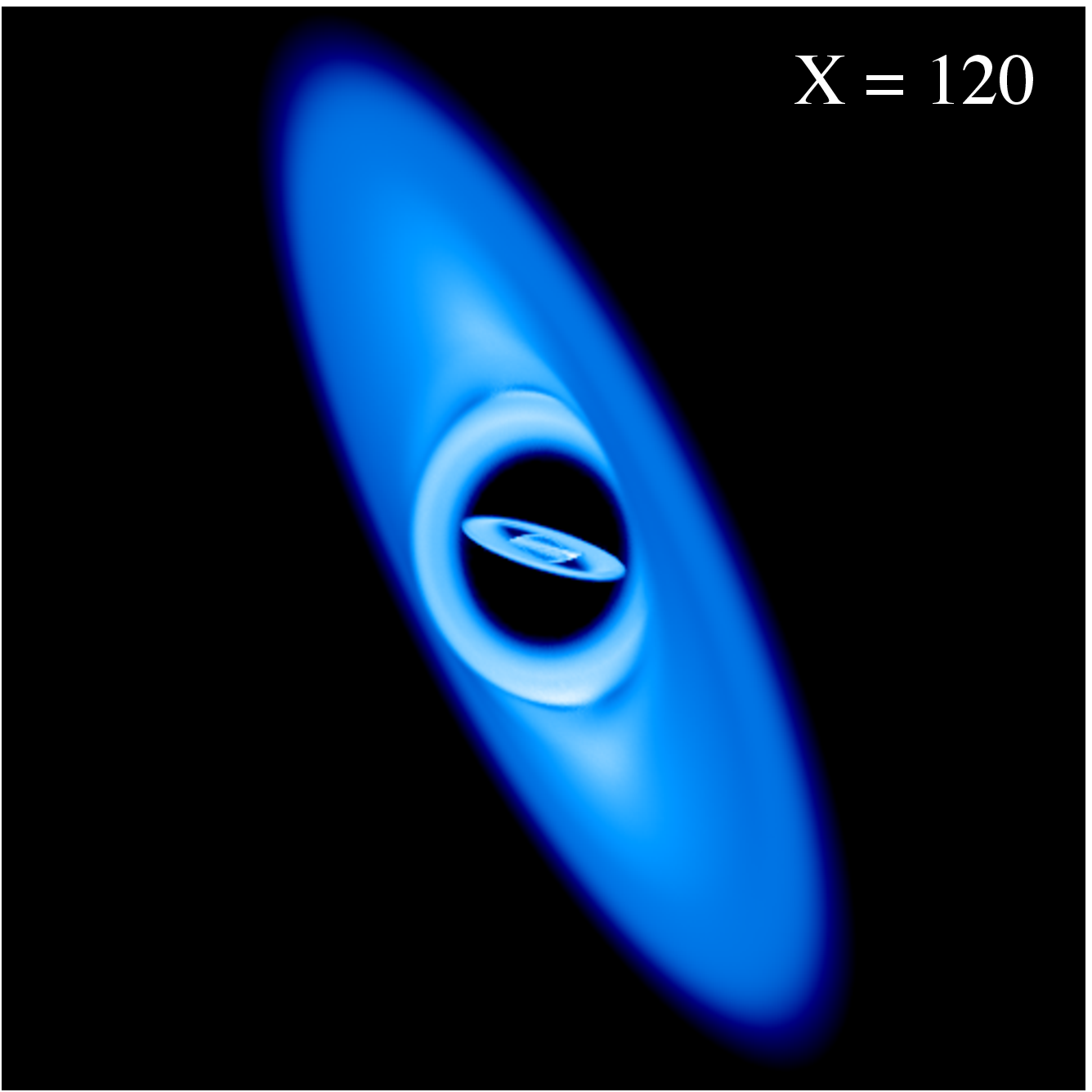}
\includegraphics[width=0.161\textwidth]{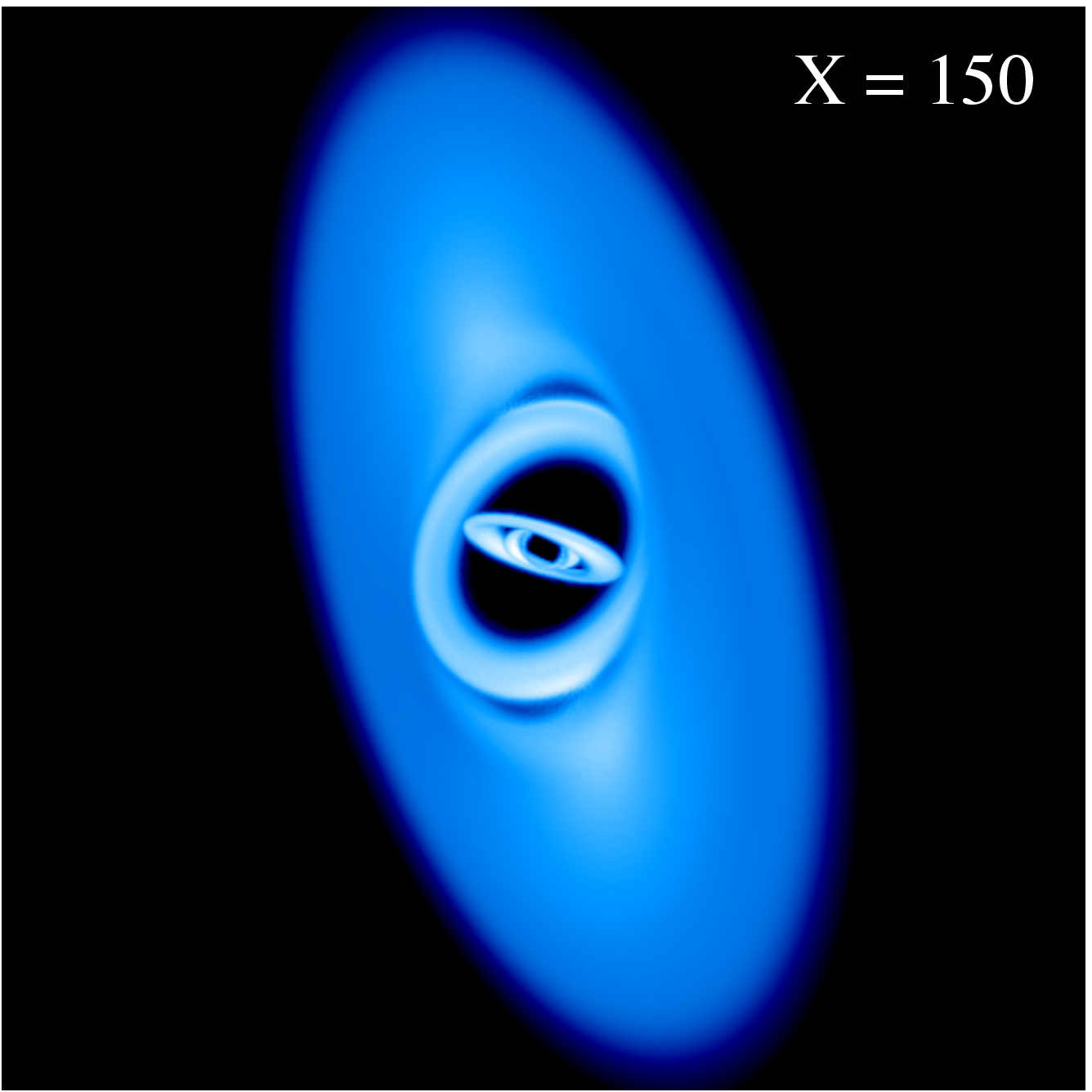}\vspace*{0.02in}
\includegraphics[width=0.161\textwidth]{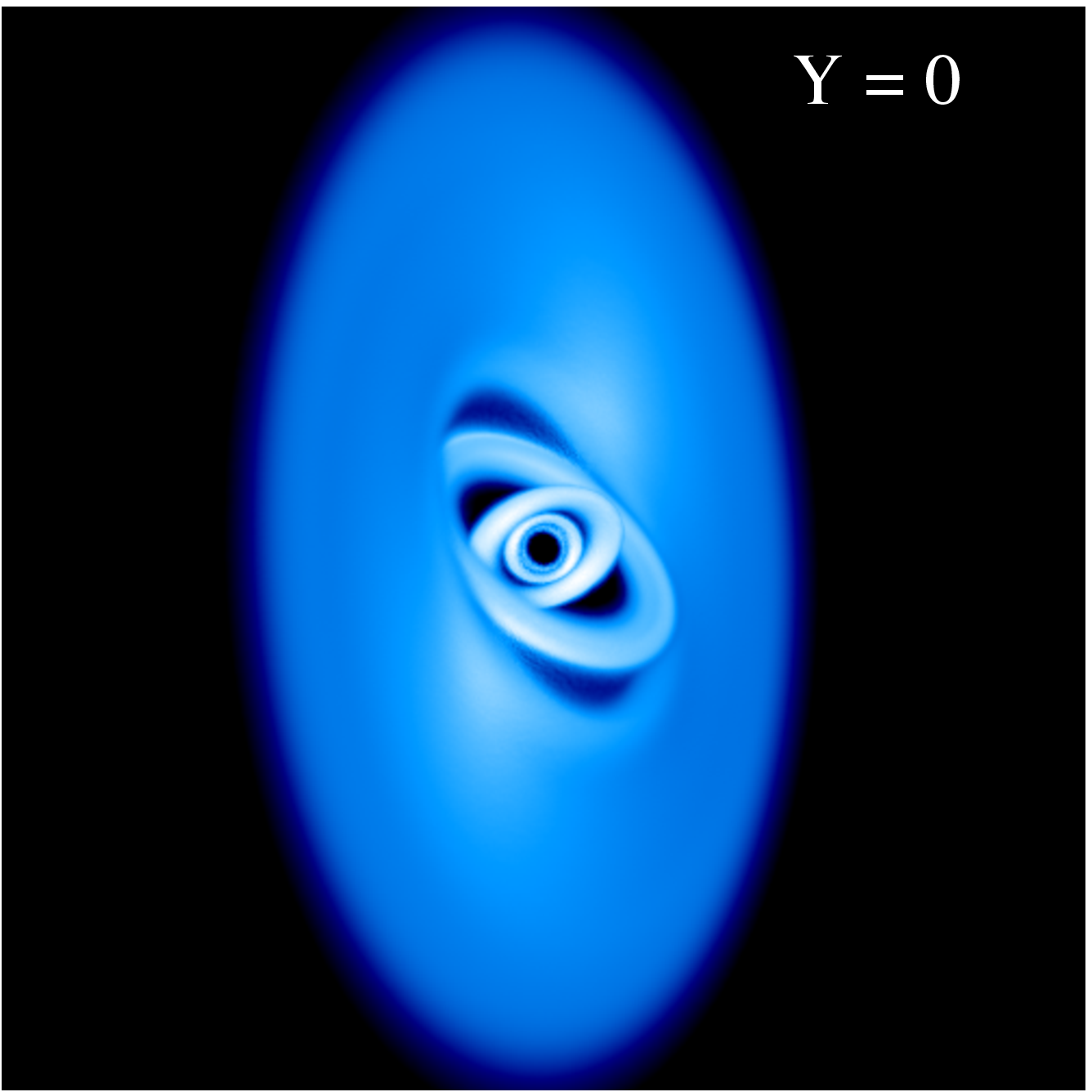}
\includegraphics[width=0.161\textwidth]{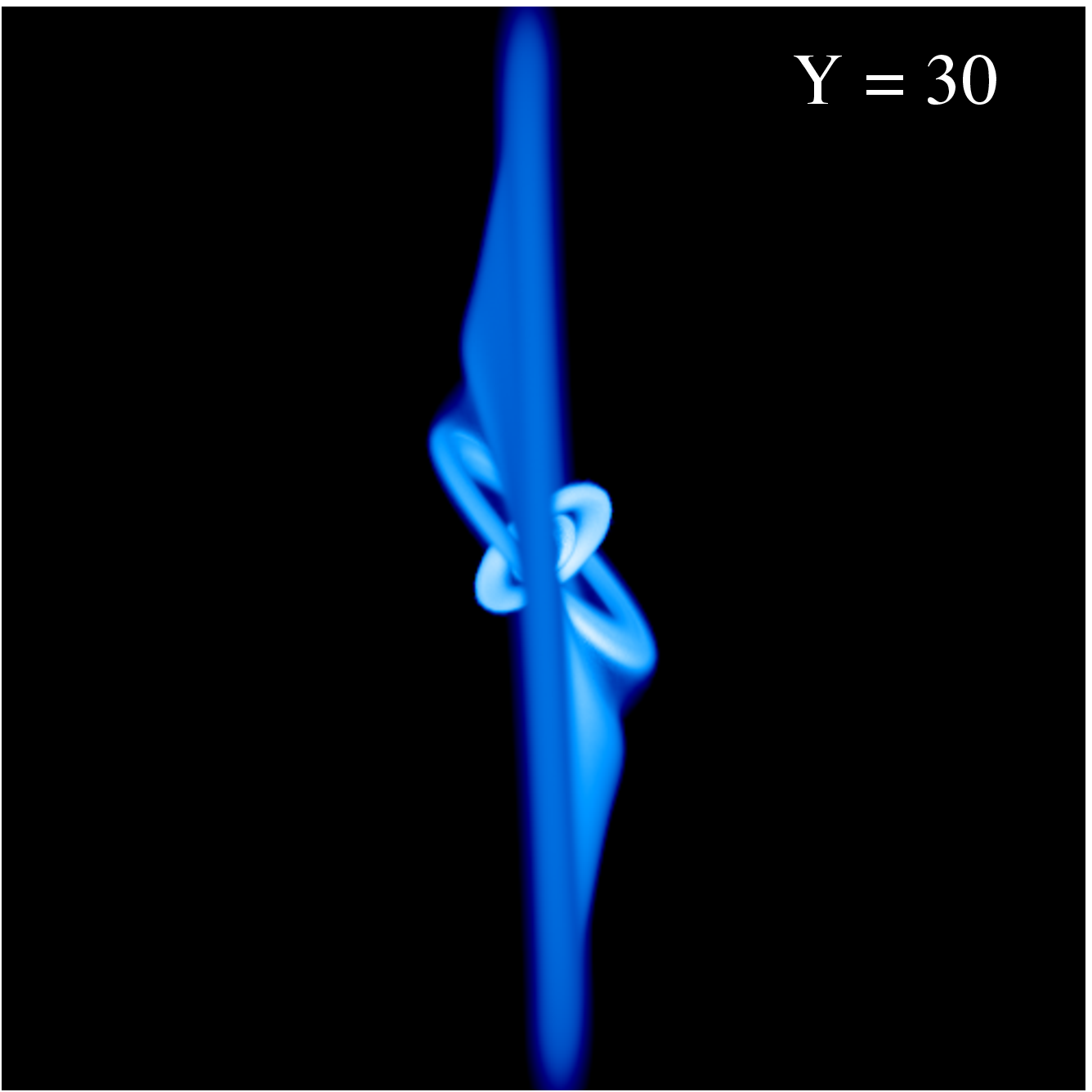}
\includegraphics[width=0.161\textwidth]{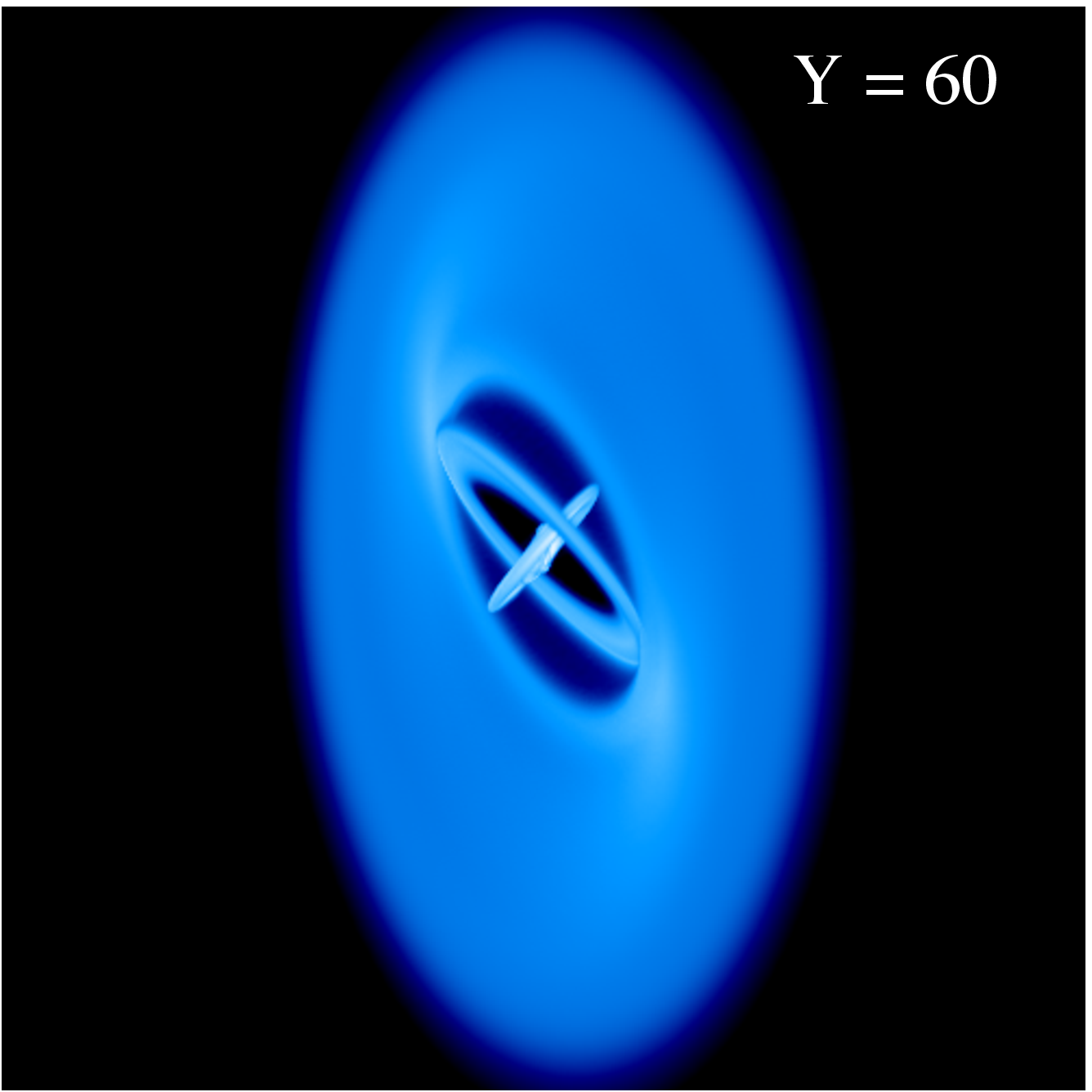}
\includegraphics[width=0.161\textwidth]{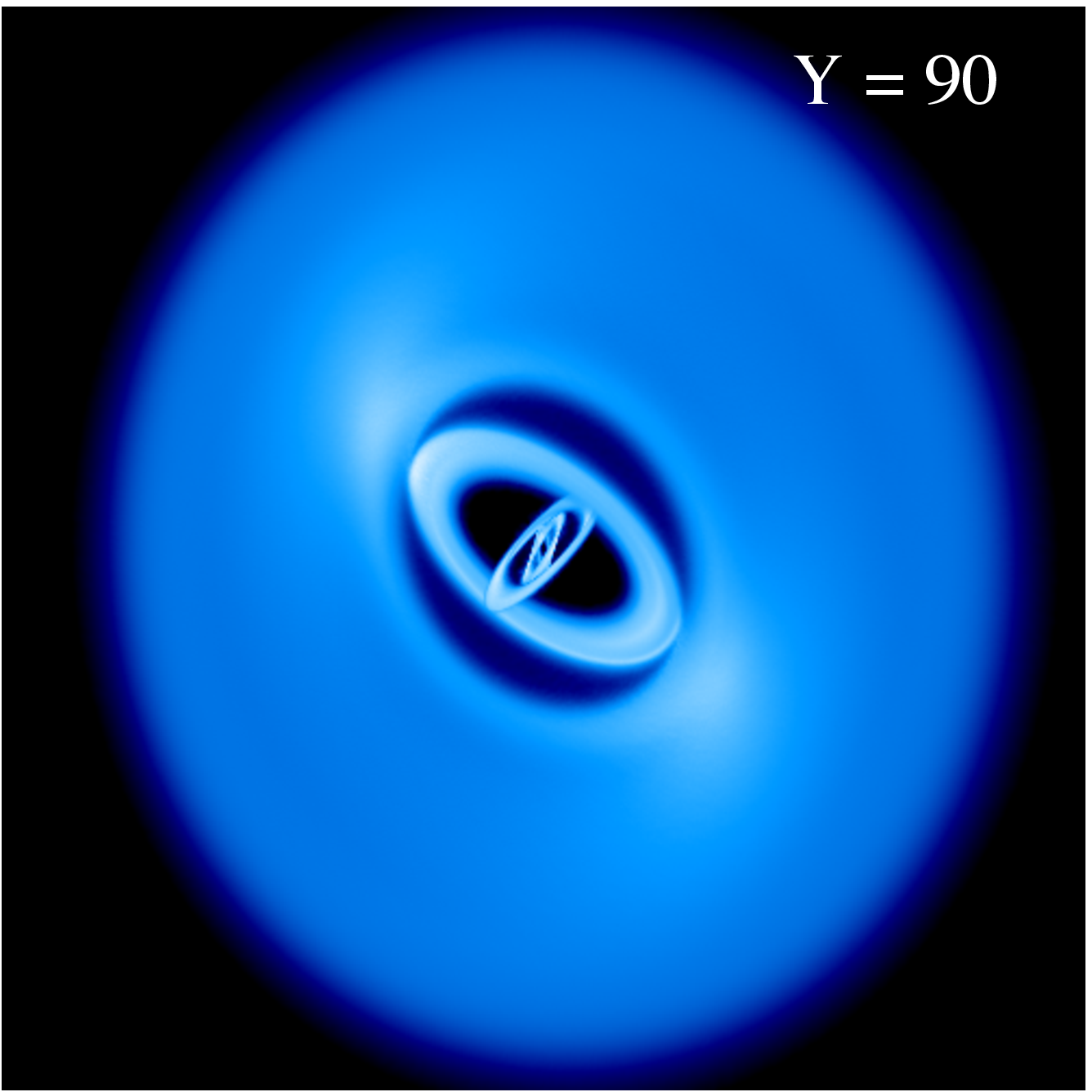}
\includegraphics[width=0.161\textwidth]{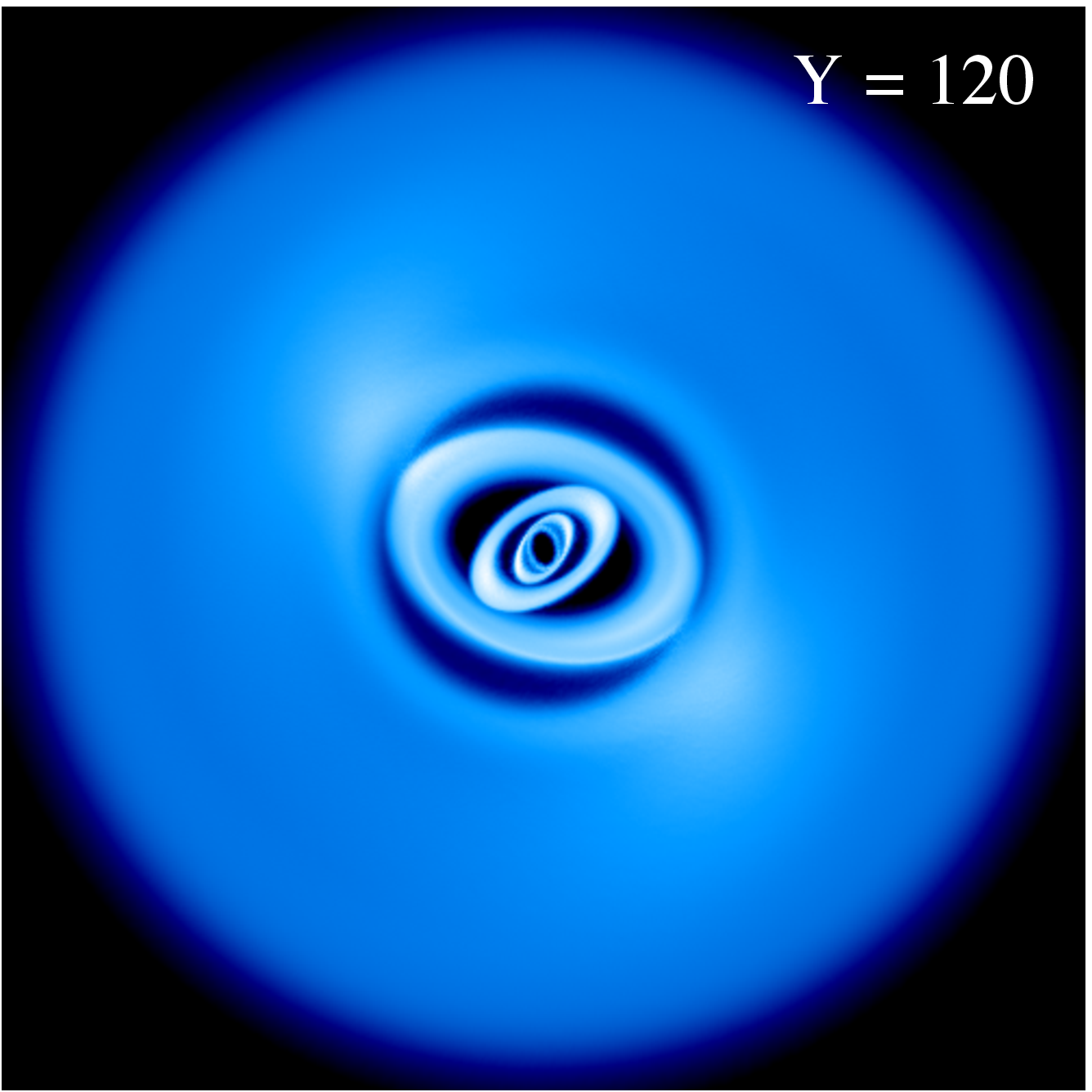}
\includegraphics[width=0.161\textwidth]{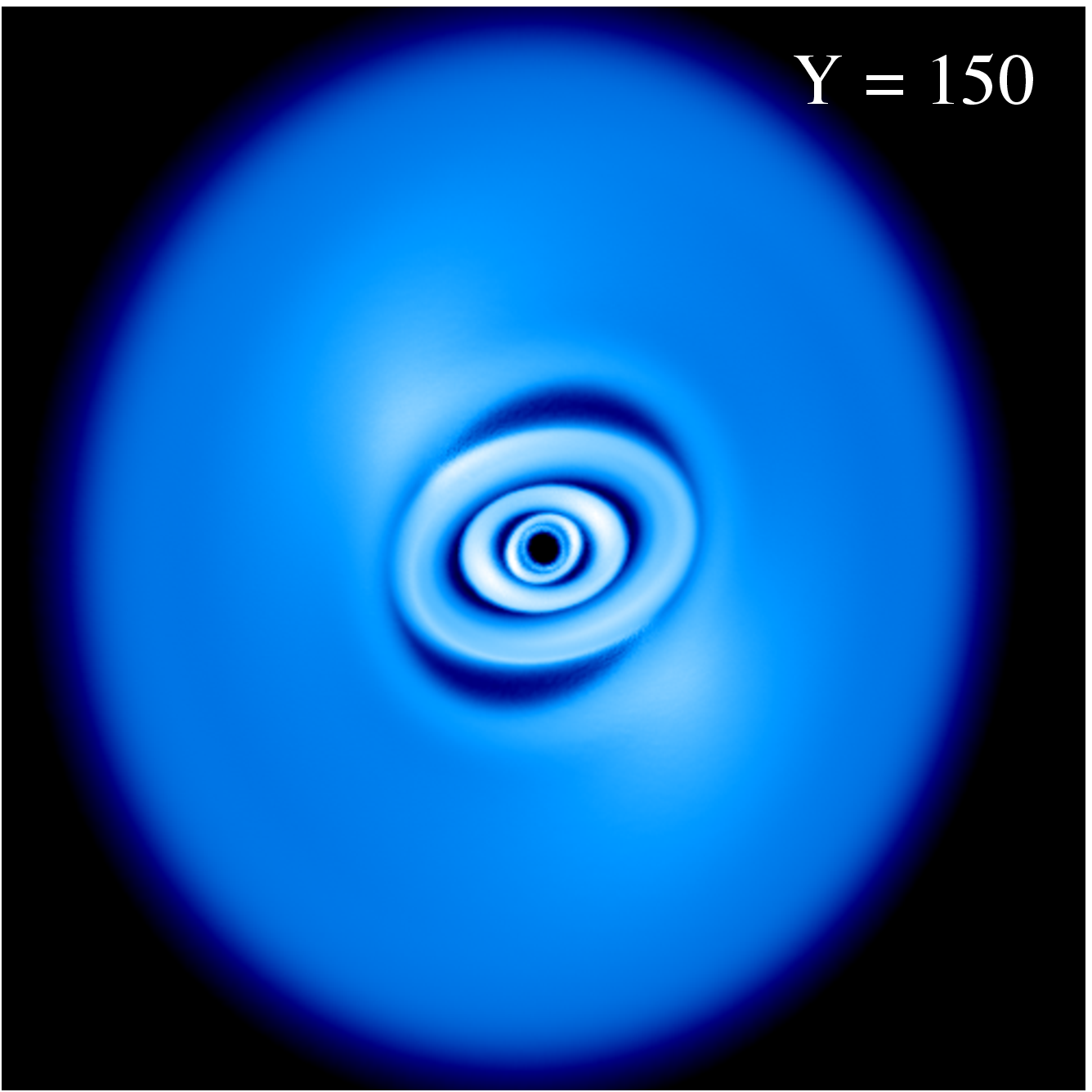}
\caption{3D renderings of the gas distribution in disc tearing simulation with $\alpha=0.03$, $H/R=0.02$ and $\theta=60^\circ$ \citep{Raj:2021aa}. Each panel depicts the same simulation at the same time, viewed from different orientation. The left-hand top and bottom row panels both show the same view---that of the $x$-$y$ plane---where the black hole spin axis, which coincides with the $z$-axis points out of the page. Then, from left to right, the top row shows the disc view with the disc rotated by an angle $X$ around the $x$-axis where the value of $X$ in each case is given in the panel. In contrast, the bottom row is rotated from the $x$-$y$ plane by an angle $Y$ around the $y$-axis. In some cases the majority of the disc is visible and pointed predominantly towards the `observer'. While in other cases, the inner disc regions are highly inclined, or obscured by either the warped outer disc or an interloping ring of matter in the unstable region.}
  \label{Fig2}
\end{figure*}

However, if the line-of-sight to the black hole passes through the disc (i.e.\@ $\theta_{\rm obs} > \theta_{\rm disc}$) then additional variability due to obscuration effects is possible. If the disc warp (i.e.\@ the outer disc regions that do not undergo disc tearing) move into (or out of) the line-of-sight then the emission properties vary significantly as the inner regions of the disc are blocked (revealed), and the timescales for such changes are long -- relying upon precession at large radii (cf. equation~\ref{tp}) or slow decay of the disc warp. However, if the interloper is instead a ring of matter in the unstable region of the disc, then we can expect the precession timescale at that radius to be imprinted on the emission as the ring blocks some of the flux from the central regions reaching the observer. The presence of multiple such rings with different inclinations blocking different parts of the disc at different times may erase a simple periodic signal, but the generic timescale for variability from a ring at a given radius is governed by equation~\ref{tp}. In Fig.~\ref{Fig2} we show an example of the effect different orientations may have on the system properties. Each panel in the figure is the same disc model ($\alpha=0.03$, $H/R=0.02$ and $\theta=60^\circ$) viewed at the same time, but from different orientations. The black hole spin axis is in the $z$-direction, and the reference views are of the $x$-$y$ plane (the top and bottom left-most panels; which are the same) and thus the black hole spin points out of the page in the left-most panels. The top row of panels, from left to right, are views of the disc starting with the $x$-$y$ plane, which are subsequently rotated by an angle $X$ around the $x$-axis. The bottom panels are the same, but now the rotation is performed by an angle $Y$ around the $y$-axis. In some configurations the central regions are clearly visible, and in others they are largely blocked from view -- either by the outer disc or a precessing ring which is crossing the line-of-sight to the disc centre.

Finally, we note that while the geometry may be important for determining which parts of the disc are seen by the observer, it is also important for determining which parts of the disc are available to be seen by the material that makes up the broad line region (typically assumed to be clouds orbiting near the outer disc regions). As the matter comprising the broad line region is typically expected to orbit close to the original disc plane, it is plausible that precessing rings in the unstable region of the disc may act to block the central disc emission from fully illuminating the broad line regions. Thus an observer looking unobstructed at the disc central regions may see a constant flux, but the flux arriving at the broad line regions may be time variable. This could cause a disconnect between the continuum and emission lines, and may thus offer a potential cause for the recently observed `broad line holidays' \citep{Goad:2016aa,Goad:2019aa}.

\section{Discussion}
\label{discussion}
\begin{figure}
    \includegraphics[width=\columnwidth]{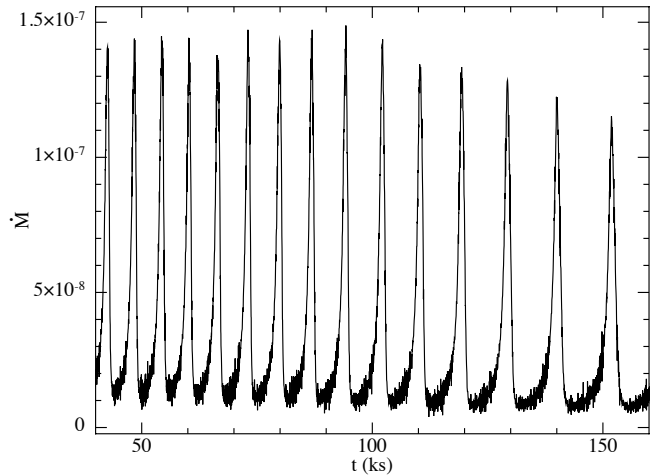}
    \caption{The accretion rate on to the black hole in the simulation presented by \cite{Raj:2021aa} with $\alpha = 0.03$, $H/R = 0.05$ and $\theta = 60^\circ$ with the disc modelled with $10^6$ particles. The time axis has been scaled such that the units are kiloseconds for a $4\times 10^5 M_\odot$ black hole. We note that the timescale on which the behaviour occurs depends on where in the disc the instability occurs (here it occurs at the inner disc edge, see Fig.~\ref{FigB} below, but this can be scaled to longer timescales if the behaviour occurs at larger radii; cf. equation~\ref{tp}). The accretion rate is in arbitrary units. In this case the disc tearing behaviour is restricted to the disc inner regions, see Fig~\ref{FigB}. The inner disc behaviour is cyclic; the innermost ring of matter is torn off, which then precesses until it interacts strongly with the neighbouring ring and is robbed of angular momentum, loses rotational support and accretes dynamically on to the black hole. The accretion rate is representative of the dissipation rate in the inner disc regions, and therefore may reflect the rate of generation of energy from the accretion flow. This should be compared to Figure 1 of \cite{Miniutti:2019aa} for GSN 069, Figure 1 of \cite{Giustini:2020aa} for RX J1301.9+2747, or for the `heartbeat' mode in X-ray binaries to (for example) Figure 1 of \cite{Zoghbi:2016aa} for GRS 1915+105.}
  \label{FigA}
\end{figure}

\begin{figure}
  \centering\includegraphics[width=\columnwidth]{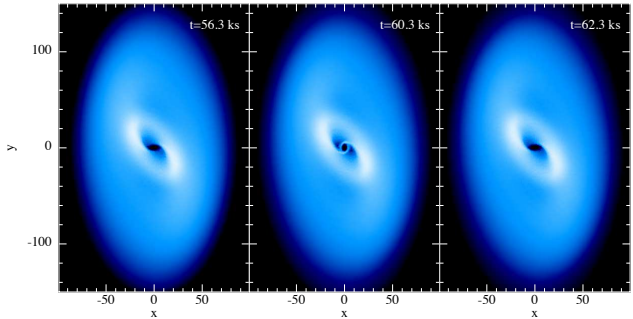}
  \caption{The disc structures from the simulation accretion rate depicted in Fig~\ref{FigA} at three times just before, during and after a peak in the accretion rate. The axes are in units of gravitational radii ($R_{\rm g} = GM/c^2$), and the colour denotes the column density with white the highest and dark blue the lowest. The times in the plots have been scaled to represent a $4\times 10^5M_\odot$ black hole (same as Fig.~\ref{FigA}). The left hand panel shows the disc in a warped state, the middle panel shows the innermost ring broken off from the disc, and the right hand panels shows a return to the warped state once the ring has been accreted.}
  \label{FigB}
\end{figure}

In the previous sections we have discussed the dynamics of disc tearing, the timescales on which variability might manifest in the observable properties, and how the disc geometry can affect what we see. Here we discuss the possible connection between disc tearing and rapid flaring variability observed in some AGN (referred to as quasi-periodic eruptions), which may also be related to the `heartbeat' modes of some X-ray binaries. In the left hand panel of Fig~\ref{FigA} we provide the accretion rate with time of the simulation reported in \cite{Raj:2021aa} with parameters $\alpha = 0.03$, $H/R = 0.05$ and $\theta=60^\circ$ performed with $10^6$ particles. In this simulation the disc tearing behaviour was restricted to the innermost regions of the disc (Fig.~\ref{FigB}). The disc inner edge warps with time, and then, once the warp amplitude is large enough, the innermost ring (with radial thickness $\Delta R \sim H$) breaks off and begins to precess. Once a large angle is reached with the next ring of the disc, shocks occur and the innermost ring is robbed of rotational support and accretes dynamically on to the black hole. This process repeats, and thus quasi-periodicity is imprinted on the accretion rate shown in Fig.~\ref{FigA}. We anticipate that this signal is imprinted on the observable properties of the system, primarily through radiating away the energy generated by the shocks between the innermost rings. Thus we expect the high-energy flux (e.g.\@ X-rays) to broadly follow the same time-dependence as the accretion rate in the simulation.

Recently \cite{Miniutti:2019aa} have reported quasi-periodic eruptions in {\it XMM-Newton} and {\it Chandra} observations of the Seyfert 2 galaxy GSN~069. The eruptions occur approximately every nine hours and are seen in observations spanning several months. The 0.4-2 keV flux increases by a factor of order 10-100 for a duration of approximately one hour. Comparison of these eruptions with the accretion rate shown in Fig.~\ref{FigA}, perhaps also with an additional X-ray emitting component to provide a base-level of flux (e.g.\@ the disc corona), are again suggestive. Applying the black hole mass of $4\times 10^5 M_\odot$ to the simulation data results in a recurrence period of $\sim 3.5$\,hr\footnote{We note that no attempt has been made to survey different disc-black hole parameters to achieve a better fit. Such an effort is not currently useful, as the black hole mass is not constrained to high enough precision. \cite{Miniutti:2019aa} estimate the uncertainty at the factor of a few level. They further estimate the black hole mass from the fundamental plane of black hole accretion \citep{Merloni:2003aa} and find a value of $2\times 10^6M_\odot$, in which case the simulated period would be $\sim 17.5$\, hr.}. Similar behaviour has also been reported for the galactic nucleus of RX J1301.9+2747 \citep{Giustini:2020aa}. Here the black hole mass is reported to be $0.8-2.8\times 10^6M_\odot$, and with the eruptions separated by $\sim 20$\,ks these eruptions are also consistent with the timescale shown in Fig.~\ref{FigA} with variations in the disc structure shown in Fig.~\ref{FigB}. We therefore suggest that these quasi periodic eruptions may be the result of the disc inner regions undergoing disc tearing, with the resulting shocks that occur when broken-off rings collide providing the additional energy dissipation to power the eruptions. This behaviour may appear as if the inner disc temperature briefly rises due to the additional dissipation of energy. If the black hole mass in GSN~069 is of the order of $4\times 10^6M_\odot$, then the nine hour eruptions are consistent with the free precession timescale (eq.~\ref{tp}) at $R \approx 25R_{\rm g}$, but due to the precession of neighbouring material (eq.~\ref{dtp}) the location is more likely to be at or close to the disc inner edge $\lesssim {\rm a~few}\,\times R_{\rm ISCO}$. It is also interesting to note that the eruptions do not seem to have been present while the source was at higher luminosity \citep[see e.g.\@][]{Shu:2018aa}. This may indicate that, as suggested by the simulations of \cite{Raj:2021aa}, variations in disc parameters (in this case the disc thickness, which is expected to be higher for higher Eddington ratio) are responsible for generating changes in the variability of the accretion flow. For the disc tearing scenario discussed here, the disc parameters are responsible for determining the location and strength of the instability and thus the timescale and amplitude of the variability.

As noted by \cite{Miniutti:2019aa}, the time-dependence is reminiscent of the `heartbeat' mode of some X-ray binaries \citep[e.g.\@ GRS~1915+105 and IGR~J17091-3624;][]{Belloni:1997aa,Altamirano:2011aa}, and recently reported in an X-ray source in the galaxy NGC 3621 \citep{Motta:2020aa}. In these systems the timescales on which the beats recur is 10-100\,s. Scaling the simulation data to a black hole of mass $10M_\odot$, the recurrence time is $\sim 0.3$\,s. In the simulation we take the black hole spin to be $a\approx 0.5$, and while a higher spin value might reduce some timescales (by moving the ISCO closer to the black hole horizon) it may also move the location of the tearing region (where the warp amplitude is highest) outwards and thus increase the precession timescale on which the periodic behaviour occurs. Similarly different values of $\alpha$ or $H/R$ would affect the location of the instability in the disc and thus the type and properties of the subsequent variability. An additional possibility in these discs is that the location of the {\it thin}, i.e.\@ radiatively efficient,  disc inner edge is truncated at a radius larger than the ISCO, with a radiatively inefficient flow inside. If disc tearing occurs at the inner edge of the thin component of the disc, then the truncation radius determines the fastest precession timescales. Increasing the radius of warping instability by a factor of 4-5 would bring the timescales into agreement for, e.g., GRS~1915+105.

\section{Conclusions}
\label{conclusions}
We have discussed the dynamics of disc tearing which \cite{Nixon:2012ad} suggest may provide a source of variability in black hole accretion, and \cite{Nixon:2014aa} discuss in the context of X-ray binaries. Disc tearing is caused by an instability of the disc warp, and occurs most prominently when the disc viscosity is weak ($\alpha\lesssim 0.1$) and the warp amplitude is large \citep{Ogilvie:2000aa,Dogan:2018aa,Dogan:2020aa}. In a companion paper, \cite{Raj:2021aa}, we present the results of numerical simulations of disc tearing around black holes. Here we have discussed the resulting dynamics, and the timescales on which we expect black hole discs to show variability. We have suggested that the large amplitude, short timescale variability exhibited by AGN may be explained by the disc undergoing such dynamics, and that this can result in intrinsic changes to the disc (both geometric and energy dissipation/accretion rate changes) and also, in some cases, to changes to the observable properties of the disc through time-dependent obscuration. These effects may account for the short timescale evolution observed in, for example, changing-look AGN. We have also shown that for some disc parameters, disc tearing can occur predominantly at the disc inner edge, and that these cases show similarity with the quasi periodic eruptions observed in GSN~069 and RX J1031.9+2747 and the `heartbeat' mode observed in some X-ray binaries. In future work, we will develop more sophisticated simulation models to connect in more detail with the available observational data.

\acknowledgments
We thank the referee for a helpful report. We thank Jim Pringle for useful comments on the manuscript. We thank Mike Goad for useful discussions on the observed properties of AGN. CJN is supported by the Science and Technology Facilities Council (grant number ST/M005917/1). CJN acknowledges funding from the European Union’s Horizon 2020 research and innovation program under the Marie Sk\l{}odowska-Curie grant agreement No 823823 (Dustbusters RISE project). This research used the ALICE High Performance Computing Facility at the University of Leicester. This work was performed using the DiRAC Data Intensive service at Leicester, operated by the University of Leicester IT Services, which forms part of the STFC DiRAC HPC Facility (\url{www.dirac.ac.uk}). The equipment was funded by BEIS capital funding via STFC capital grants ST/K000373/1 and ST/R002363/1 and STFC DiRAC Operations grant ST/R001014/1. DiRAC is part of the National e-Infrastructure. We used {\sc splash} \citep{Price:2007aa} for the figures.

\bibliographystyle{aasjournal}
\bibliography{nixon}


\end{document}